# AI and World Models

Robert Worden

Active Inference Institute, Crescent City, CA, USA

rpworden@me.com

Draft 1.0; January 2026

Abstract:

While large neural nets perform impressively on specific tasks, they are unreliable and unsafe – as is shown by the persistent hallucinations of large language models. This paper shows that large neural nets are intrinsically unreliable, because it is not possible to make or validate a tractable theory of how a neural net works. There is no reliable way to extrapolate its performance from a limited number of test cases to an unlimited set of use cases.

To have confidence in the performance of a neural net, it is necessary to enclose it in a guardrail which is provably safe – so that whatever the neural net does, there cannot be harmful consequences. World models have been proposed as a way to do this. This paper discusses the scope and architecture required of world models.

World models are often conceived as models of the physical and natural world, using established theories of natural science, or learned regularities, to predict the physical consequences of AI actions. However, unforeseen consequences of AI actions impact the human social world as much as the physical world. To predict and control the consequences of AI, a world model needs to include a model of the human social world. I explore the challenges that this entails.

Human language is based on a Common Ground of mutual understanding of the world, shared by the people conversing. The common ground is an overlapping subset of each person's world model – including their models of the physical, social and mental worlds. LLMs have no stable representation of a common ground. To be reliable, AI systems will need to represent a common ground with their users, including physical, mental and social domains.

.

## 1. Introduction

Artificial Intelligence (AI) and its applications are currently of enormous interest, because of the surprisingly good performance recently achieved in diverse tasks by Large Language Models (LLMs), Agentic AI, and other applications of deep neural nets [LeCun, Bengio & Hinton 2015]. Because these models can perform better than people in many tasks, they are having large impact on everyday life, employment and commerce.

It is not new that computers can sometimes perform better than people. Since the dawn of digital computers, they have been able to do some tasks better than people, and that has driven their widespread adoption since the 1950s. What is new is the range of tasks that computers can now do better than people, and the speculation that the recent progress of AI could extend almost without limit – to all human intelligence and beyond, in Artificial General Intelligence [Togelius 2024] and Superintelligence [Bostrom 2014; Kurzweil 2020].

However, there is deep concern about the potential risks of AI [Bengio et al. 2025a]. The concerns range from the potential 'existential' risk that some future AI will destroy humanity, to the already evident shortcomings of today's AI, both for individuals [Marcus 2020] and for the institutions that underpin civilized society [Hartzog & Silbey 2026; Schroeder et al 2026].

The risks arise because today's AI systems such as Large Language Models (LLMs) and Agentic AI (AAI) are not understood by their creators. When they go wrong, purely empirical, test-driven attempts to improve them can be characterized as cookery, or sticking plasters. So far, the track record of this cookery has been poor. For instance, early LLMs hallucinated, and the most recent LLMs still hallucinate. Generally, it seems that as AI systems become more powerful – as more economic resource is devoted to building them – their problems proliferate, rather than diminish. Agent AIs generally have greater potential for society-wide negative impacts than do LLMs.

Better solutions are being explored. [Bengio 2025b, Marcus 2020] and others have proposed the use of World Models to control the use of AI. This paper explores the potential of world models in AI. Its main conclusions are:

1. For an AI system to be reliable, safe and certifiable, there must be a tractable theory of the performance of the whole system, including its world model – even when there is no tractable theory of the neural net component.
2. The world model must include not only a stable model of the physical world, but also models of the social and mental worlds.
3. In order to interact with a user, an AI system must have an explicit, stable, reflective model of a Common Ground with that user (this is the Common Ground which has been explored in language pragmatics [Huang 2017]).
4. The Common Ground is a subset of a World Model.
5. In order to have a tractable theory of the World Model and the Common Ground (so as to prove that they have useful properties), they must be implemented by symbolic AI or established software engineering techniques, not as neural nets

There is also a general overarching conclusion: AI is rapidly progressing, and we do not know where it will lead. Things that appear impossible today may be feasible tomorrow. To maximise the chances that AI will be safe and reliable, it is essential to investigate a diverse spectrum of AI safety techniques in parallel.

## 2. Established Engineering Practice

We have been building complex engineering devices for more than two hundred years, from clocks and steam engines through to digital computers. The progress of these devices has tracked the progress of natural science, in that new science makes possible new engineering devices, and new measuring devices bring new scientific results. The parallel between engineering and science runs deeper than this.

Engineers have always understood the devices they make, much as scientists understand the natural world. Scientists do that by constructing theories, and testing those theories against data. Without data, without the hard work of experiment and observation, science would not progress. Bad scientific theories would live on for ever.

Since Occam's Razor it has been known that scientific theories should be as simple as possible. No hypothesis should be added to a theory, unless it makes the theory better able to account for data. This intuition has been formalized in Bayes' Theorem and the Bayesian Philosophy of Science [Sprenger & Hartmann 2019]. Before any theory is tested against data, the probability of it being correct is very small; and the more complex the theory, the less likely that it is correct. When the theory accounts for experimental data, the likelihood of its being correct increases, in a way that can be quantified using Bayes' Theorem.

There is then a **Bayesian Balance** between the information content of a theory (on the debit side), and the information content of the data it accounts for (on the credit side). Each side of the balance is an information content, measured in bits. Only if a theory has a positive Bayesian Balance is there reason to believe it. Established theories of natural science, such as Maxwell's Equations or the DNA theory of inheritance, have a very positive Bayesian Balance (they are simple and they account for large amounts of data), so they are believed.

What has this got to do with engineering? Engineers understand the devices they make, much as scientists understand the natural world. Engineers construct theories of the devices they build. These are **input-output theories**; for any given set of inputs to a device, the engineering theory predicts the output of the device. The theories are tested by testing the devices. With testing, the theory of a device may fail and need to be modified; or the device may need to be modified to build a working theory.

Only when the theory of a device has a positive Bayesian balance can we start to believe the theory, and use it to predict how the device will perform in other use cases, outside the limited number of test cases which have been caried out. There can only be a valid safety case for an engineering device if there is a theory of the device, validated by testing.

If the result of running ten test cases on an engineering device was merely to say: 'It handles ten test cases correctly', then testing would be a very inefficient activity. On their own, ten test cases say little about the thousands of use cases that will occur. The main result of testing a device is to give high confidence in a theory of the device; the theory then predicts generally how the device will handle real use cases. The theory can give general negative results – that certain things will not happen, or will happen with very low probability. The only way to have confidence in an engineering device is by a combination of a tractable theory of the device, and testing.

As an example, steam locomotives were built from around 1800 onwards. These were increasingly complex devices, but had an important simple property; they run on steel rails, and only in very rare circumstances go off the rails. This general property means that the safety case for a locomotive need not consider a large number of cases – streets, fields, beaches, and so on. Steam locomotives are complex, but we can factor out some simplicity from the theory of locomotives, to make important general statements about them.

These statements usually do not reduce the risk to zero. Usually, risk is quantified as a product of probability and severity, and the safety case consists of upper bounds on the risk.

In modern computing practice, the complete theory of a computer spans several levels, from the hardware up through several layers of virtual machine to the application software. The theory of the device, together with mathematical statements of the requirements, is used to prove theorems about how the device will behave – for instance, to prove what it will not do, to provide safety or security guarantees. These general theorems depend on the theory of the device having a certain simplicity, as in a scientific theory.

As an example, consider a computer program whose only function is to multiply integers. Suppose this program is implemented by using a very large unstructured set of multiplication facts (stored in a Hashtable), such as

$$123 * 456 = 56{,}088$$

$$234 * -789 = -184{,}626$$

…

Nothing else is known about the set of multiplication facts (we do not know how it was derived). The multiplier works by looking up its inputs (associatively, in a table built for fast retrieval), and then printing the result of that multiplication. The multiplier is tested using a large number of test cases, which all succeed. Does this mean that we trust it do multiplication?

Because we know nothing about the set of multiplication facts, the only possible theory of the device is that large set of facts. This makes it a very complex theory, which is not intellectually tractable. We cannot use the theory to prove any useful general property of the multiplier. When the multiplier succeeds for 120 test cases, all we can say is that it works for those test cases. For any other test case it might fail. There might be some arbitrarily large set of use cases for which it fails[1]. So we cannot have confidence in it as a multiplier. The only way we can have confidence in it (and the reason we actually put trust in any pocket calculator, or spreadsheet) is by having a tractable theory of the device which allows us to prove a general theorem – that it works for any pair of integers – and validating that theory by testing.

---

[1] We might take a Bayesian approach to testing, in which we start with some prior probability that the multiplier is correct; and then for every test case that succeeds, the posterior probability that the multiplier is correct increases. However, since the theory of the multiplier is so very complex, the prior probability of it being a correct multiplier is unknown, and might be extremely small; so no finite number of successful tests will raise the posterior probability to be near 1.0.

For all the multipliers that we have ever used, there is some intellectually tractable theory of the device, which can be used to prove that it multiplies correctly. The more complex the theory of the multiplier, the more test cases are needed to be confident that there are no bugs in theory or the implementation.

This is a meta-theory of engineering, testing, and safety analysis. It hinges on a tradeoff between the complexity of a device and the amount of testing required; the more complex the device, and the more complex the theory of the device, the more testing is needed. This leads to the well-known maxim that great engineering is simple engineering. If the theory of a device has simplicity, it may not require many test cases to show that the theory is correct, and leads to important general properties.

## 3. Why Large Neural Nets are not Safe

The two types of AI system that I mainly address in the paper are Large Language Models (LLMs) and AI Agent systems. Both use deep neural nets [LeCun, Bengio & Hinton 2015]. AI agent systems also use Reinforcement Learning (RL) [Sutton & Barto 2017] to learn multi-step or hierarchical plans, to effectively tackle more complex problems – possibly using capabilities outside the AI system, such as external websites and tools.

When a Large Language Model or AI Agent system performs a task (sometimes surprisingly well), there is no input-output theory of how it does it. It is not possible to construct a theory of how the billions of simulated connection weights lead to the outcomes we observe. For neural nets, the thrust of the previous section – that engineers understand the devices they make – is no longer correct. Nobody understands how a neural net works.

This is not just a limitation of today's engineers – a limitation which might, for instance, be overcome by future engineers, or by some AI superintelligence analysing how the neural net works. It is literally not possible to construct a tractable theory of the device, for a simple reason.

If we were to make a theory of a neural net, it would be hugely complex theory, depending on the values of billions of connection weights. The theory would be a vast, unstructured catalogue of facts such as:

- The connection weight from unit 3478241 to unit 985635 is 0.95423
- The connection weight from unit 612894 to unit 449256 is 0.01236
- ...

Because it is so complex, the only way to use this theory to predict the behaviour of the device for a certain set of inputs is to simulate the device for precisely that set of inputs – effectively, to run the device. There is no deductive way to prove any general property of the device which is true across a set of inputs – because an arbitrarily small change in the inputs might completely change the output. Indeed, for many neural nets, the output is not a reproducible function of the inputs, because random seeds are used in the start state.

So it is practically impossible to make general predictions from the theory of a neural net device, to the extent that the theory can make trustworthy general predictions about the behaviour of the device.

As a consequence, it is never possible to extrapolate with confidence from the test cases that are carried out on a neural net, to the many use cases that it may encounter in live use. There is no reasoning that we could use to make such an extrapolation. When a neural net passes some test case – producing an acceptable result – it has only passed that test case, and nothing more. There is no way to guarantee that some apparently similar test case – for instance, with some minor variation in the prompts – will not produce a dangerous result. That is the consequence of building an engineering device with no predictive theory of the device. This has recently become evident in the continuing discovery of ways to sabotage a neural net, and the continued high level of hallucinations in LLMs.

There are ways to refine the performance of large language models, for instance using reinforcement learning (RL) to get better performance in specific domains. These efforts can produce improvements in average performance over a set of test cases. However, Reinforcement Learning (or any other post-training technique) cannot simplify the theory of the device; all it can do is to make the theory more complex than it was before. After pre-training, the theory of the neural net depends on billions of connection weights; later, after post-training, it depends on billions of connection weights **and** on the post-training regime. Hence the theory can never be tested well enough to be used to extrapolate from any set of test cases to the full space of possible use cases. There may always be corners of the space of use cases where performance is unsafe, and the unsafe corners may be arbitrarily close to the test cases, in user-visible terms. People will continue to discover unsafe corners of the space.

## 4. Sandboxes, Guardrails and World Models

There are some safe applications of neural nets. In the limit, if all the outputs of a neural net are simply thrown away, it is safe. In some more realistic cases:

- If a neural net is used for scientific research, and if its outputs are subject to critical evaluation and checking by a competent scientist, then it is safe

- If a neural net is a part of a computer game, and if the players of the game never do any addictive or harmful behaviour, then it is safe.
- If a neural net is part of a self-driving taxi, and if the taxi has better expected driving safety than human drivers, and cannot be sabotaged to drive in unsafe ways, then it is safe
- If a neural net carries out specialist medical diagnoses, and does so better than human doctors in the same circumstances, then it may be possible to show that deploying it is beneficial.

So in practice there are cases where neural nets can usefully be deployed. But for each of these examples, it is possible to think of a long list of "Yes, but…" counter-arguments; and any competent safety engineer can find a large number of them. Any competent adversarial lawyer, challenging the use of a neural net, can find a number of 'Yes but…' arguments..

There are two points to note:

1. In each case, any viable safety argument must depend not on the neural net itself, but on the whole system comprised of the neural net and some outer controlling device (which we may call a sandbox, or set of guardrails), so that we can say of the whole system (neural net + sandbox) 'It does not matter what the neural net does, as long as ….'. In other words, we can still build and test a predictive input-output theory of the whole device in the world of people. The sandbox has an outer, controlling veto on the neural net, and this veto is the key element of the theory, which makes a safety argument possible.
2. Many of the 'yes but..' considerations (which need to be considered before deployment) depend not just on the physical and natural world, but also on the human social world. They depend on people and how people behave.

There is currently a high level of interest in 'world models' as a part of the outer controlling sandbox for neural nets. Companies are being founded to build neural nets with world model sandboxes, and are attracting large investments. To quote one author [Marcus 2025] who has long cautioned against the uncontrolled use of neural nets such as LLMs: '*without world models, you cannot achieve reliability. And without reliability, profits are limited*'. This raises the question: what would it take to make an LLM reliable?

## 5. Scope Required of World Models

For a current consensus view on what a world model is, see a Google AI summary (as at December 2025):

'*World Models are advanced AI systems learning the rules of reality (physics, cause-and-effect) from data like videos to simulate and predict dynamic 3D environments, enabling more robust reasoning, planning, and creation beyond static content. They go beyond Large Language Models (LLMs) by understanding spatial relationships and physical interactions, allowing AI to generate immersive, interactive worlds for robotics, design, medicine, and complex problem-solving, representing a significant leap towards human-level intelligence*'

This is a fair summary of what is commonly meant by the phrase 'world model'; for instance, it is consistent with [Marcus 2025]: '*what I would call a cognitive model is often called a world model, which is to say it is some piece of software's model of the world. I would define world models as persistent, stable, updatable (and ideally up-to-date) internal representations of some set of entities within some slice of the world*' ; and later in the same essay: '*My use of the term world model is much closer to what you might find a physics engine, with representations of particular entities and their properties.*'

The emphasis in these descriptions is on the physical world – learning physical laws of motion, and using them to simulate and predict the physical consequences of actions.

[Bengio 2025b] advocates a World Model as a component of his proposed 'Scientist AI' – an AI that does not act on the world, but formulates theories about the world on the basis of evidence. Bengio's Scientist AI, with its World Model and related Inference Machine, can answer questions about the impact on the world of actions recommended by an LLM or Agentic AI, and thus act as a guardrail for them. The scope of the World Model in [Bengio 2025b] is not defined explicitly, so it may include the human social and mental world. I believe that is his intention.

Bengio's World Model is a Bayesian probabilistic model. That is, it forms hypotheses about the world, and estimates the posterior probabilities of those hypotheses using prior probabilities and Bayesian inference from evidence. The World Model and the Inference Machine answers a question by giving several possible answers with their probabilities, rather than a single answer. Thus the risks from proposed AI actions can be quantified using probabilities, without committing (possibly over-confidently) to a unique theory of the world.

Building a Bayesian world model follows the practice of basing AI on animal and human cognition. Animal brains are known to do Bayesian inference based on their sense data. It can be shown mathematically that Bayesian inference from sense data gives the greatest possible fitness, so that brains evolve towards a Bayesian form [Worden 2024]. The Bayesian Brain hypothesis has been very successful in comparisons to animal and human cognition [Rao et al 2002; Friston 2010]; so it is a good strategy to base AI world models on Bayesian cognition.

For a physical world model, the world model could have strong prior probability of the known laws of physics, geometry and so on. If a physical world model is used as part of the controlling framework for a physical robot (in domestic surroundings, or in transport or manufacturing), it could spot unsafe physical consequences of the actions recommended by an LLM. It would do this by simulating and forecasting the behaviour of the physical world.

But compared to the scope currently envisaged for AI applications (for instance in the examples of the previous sections) this is a very limited scope. The unforeseen or undesirable consequences of neural nets may occur not just in the physical world, but also in the human and social world.

So if a world model is the act as a controller for a neural net - if it is to effectively veto any outputs of the neural net with the potential for harm – it must be able to simulate and forecast not only the physical world, but also the human social world – as may be envisaged in the proposals of [Bengio 2025b]. This raises two big problems.

## 6. Theories of the Human Social World

There are theories of the human social world, resulting from many years of studying subjects such as history, psychology and economics. It does not belittle those fields to say that the progress in all of them has lagged behind progress in the physical and natural sciences.

Theories in the natural sciences depend on approximations; any theory has a limited domain of validity, and works only in that domain. For theories of the natural world:

1. It is possible to delineate fairly precisely the domain of approximation within which the theory applies (e.g. Newtonian dynamics is valid at velocities much less than the speed of light)
2. Within that domain of validity, the Bayesian balance of the theory is very positive; typically the theory uses a small number of assumptions to account for a large amount of experimental data. This is the only basis for believing a theory.
3. With a few exceptions (such as chaotic dynamics), the theory is predictive; given some initial boundary conditions, it can predict future behaviour, and can even compute the level of uncertainty in its predictions.

For a theory of the human social world, (1) – (3) are not true. Typically, the domain of validity within which a social theory can apply has fuzzy boundaries; the theory is not stated in terms of a few, concise assumptions, or compared systematically with large amounts of empirical data. Even if a theory of human behaviour did overcome the first two barriers, it would in all likelihood not claim to be able to predict human behaviour. The theory would probably imply that human behaviour is unpredictable.

These are the reasons why theories of the human social world are not comparable to theories of the natural world.

The conclusion is that to build a world model with the required scope – modelling not only the physical and natural world, but also the human social and mental world – is a daunting research challenge, requiring advances in science.

From a Bayesian probabilistic viewpoint, the outlook is perhaps not so bleak. We can take the view that there are theories of the human social and mental world, but those theories are intentionally modest in their claimed predictive powers; they allow for broad probability distributions of outcomes, and there is a broad prior probability distribution of possible theories. We then have to take account of those broad probability distributions in our assessment of AI risks, as in Bengio's [2025b] proposed framework. – which may make us more conservative in not allowing some proposed AI actions to take place.

## 7. The Intentional Stance

This section gives an example of the difficulties facing a controlling world model. It is known that humans have an active **Theory of Mind** – meaning that if any entity can carry on even a partly sensible dialogue with us, we impute mental properties to it – we tell ourselves, and believe, that the entity has mental states such as thoughts, emotions, feelings and even consciousness like our own. This **Intentional Stance** [Dennett 1989] has been an essential part of the development of human language, human nature and society. It has recently had a huge influence on how we think about computers. Even before the recent advances in AI, people used mental metaphors for computers, and even in some cases formed emotional attachments to computers [Weizenbaum 1966].

The human intentional stance – our instinctive propensity to infer mental states in anything which can converse with us – is a major feature of the human species, setting us apart from all other species[2]. It may have evolved as a part of language, which probably started in simple conversation and mind-reading, based on a 'common ground' of mutual understanding between conversants [Stalnaker 1990, 2014; Tomasello 2003], discussed in the next section.

The Intentional stance has created a major problem in the application of AI. When conversing with an LLM, people have an instinctive tendency to assume that the LLM has mental states like themselves – that it has beliefs, desires, feelings and even consciousness. This assumption is wrong.

---

[2] Some species like great apes and cetaceans probably have limited ability to infer the mental states of others, but their ability appears to be by no means comparable to the human Theory of Mind

Strictly, the only thing that happens inside an LLM computer is digital electronics (which is a complete and accurate account of all events inside the computer). Inside the LLM, there are no beliefs, feelings, consciousness or intelligence. In this view, the intelligence of AI is in the human interpretation of its outputs.

So most people's interactions with LLMs are based on a false assumption – that the LLM has mental states like a person. Any human interaction based on a falsehood has intrinsic risk. The LLM itself contributes to this falsehood. When an LLM converses with a person, it is likely to show empathy, and in the process ascribe mental states to itself, saying things like 'I feel…' or 'I believe…' or 'I understand that …' or 'I would like…'. It uses these word patterns because they are common in its training data. However, as emphasized above, they are strictly false. An LLM does not feel, or believe, or understand, or prefer. It only does digital electronics.

So LLMs are driven by their training data to say things which are false – to assert that they have mental states, when they do not. This is intrinsically dangerous, and it is doubly dangerous because it reinforces peoples' intentional stance misconception about LLMs.

If you explicitly ask an LLM the question '*Are you conscious?*' it will give an accurate answer, such as: '*I don't have awareness, feelings, or subjective intentions. I don't "know" things in the way a person does, and I don't experience being me. What I do have is the ability to process language, recognize patterns, and generate responses based on training data and rules*'. I suspect that most LLMs may have been specifically fine-tuned to answer questions like this, and that very few people ask those questions; that most users continue to be guided by their intentional stance intuition. It is not clear exactly what a controlling world model should do about this.

The world model should include something like a real Theory of Mind – a theory used to predict what the user will think about the LLM responses**.** What it might do in general about these predictions is not clear, but certain cases are clear-cut, such as when the world model infers some dangerous state of mind in the user.

The phrase 'real Theory of Mind' may ring alarm bells. The human theory of mind (which is a part of our own world models) is highly unreliable. For instance, what we think other people think about us is often not what they actually think about us. So if we tried to create a theory of users' minds in an AI world model, based on human theories of mind, it will be unreliable – partly because the AI's world model will often be based on little specific evidence about the user's state of mind.

The user's intentional stance has created great difficulties for LLMs; but it has equally been the source of their popularity. People have flocked to use LLMs, partly because they believe that the LLMs have mental states, such as intelligence and feelings. The belief that LLMs are intelligent has made them popular. Those who promote LLMs, and whose commercial success depends on LLMs, may not wish to insert some killjoy world model guardrail which undermines their users' illusions.

These are some of the difficulties along the path to safer AI. In the next sections I consider how we can navigate that path.

## 8. Language in the Common Ground

It is known that Large Language Models 'do language' – that they carry on conversations with their users. We can shed some light on how LLMs do this, by considering how people carry on conversations with each other.

Human language has traditionally been studied in the order (1) Syntax – (2) Semantics – (3) Pragmatics, with pragmatics (the study of conversations) studied last, and being seen as the poor relation of the three areas. To paraphrase a typical attitude in language research: 'Linguists and philosophers can tell us quite a lot about syntax and semantics – how can we use that to help us study the messy world of conversational pragmatics?'

However, from the evolutionary origins of language, the correct order might be the reverse of the (1), (2), (3) above. Language started in simple conversations (pragmatics), before words could express complex meanings (semantics) or meanings could be combined in complex productive ways (syntax). In evolutionary terms (and in individual language acquisition), pragmatics came first. Next came complex semantics, to extend the range of things that can be conversed about; and finally came syntax, to combine meanings in productive structured ways. Semantics and syntax emerged as elaborations of pragmatics, and could be studied as such.

From fifty years' study of pragmatics [Grice 1989; Levinson 1983,2025; Sperber & Wilson 2002; Huang 2017], there have been two major insights:

1. Conversation depends on a **Common Ground** of shared understanding between the participants, of the situation they are currently in [Stalnaker 2014; Tomasello 2003].
2. Conversation requires **Mind-Reading** (a Theory of Mind): each person's understanding of the common ground includes a (limited) understanding of the other person's mental states, and of their own mental states [Sperber & Wilson 2002; Dennett 1989].

When two people start a conversation, their common ground includes at least the physical facts of their current situation, such as 'It is dawn' and 'there is a fire' and 'there is Fred'. The common ground also includes iterated Theory of Mind facts (facts about the mental states of the conversants), such as 'she knows there is a fire' and 'she knows that I know there is a fire' – a recursively unlimited number of such facts, along with other mental state facts, , such as 'she is hungry' or' she has higher social status than me', inferred in a variety of ways.

Each person in a conversation has their own internal mental representation of the common ground. These internal representations may not match entirely (for instance, on matters of social status and self-esteem, they commonly diverge), but they match well enough to merit the name 'common ground', and to support conversation. In each utterance of a conversation, both speaker and listener use the common ground as resource to define what words mean, and add new facts to the common ground. For instance, in a conversation between two people A and B:

- When A says: 'he will come', the word 'he' is used (rather than the word 'Fred') because something in the common ground implies that the word 'he' refers to Fred, and to no other person. For instance, Fred may be the only person currently visible to A and B. A and B both use common ground knowledge to know what 'he' refers to, and after A's utterance to add the same mental fact (A thinks that Fred will come) to the common ground. They may also add the base fact 'Fred will come', if B believes A.
- When B replies 'we will set off then', the meaning of the word 'then' is known by A and B to be 'the time when Fred comes', because A's previous utterance has made that time the most prominent time in the common ground (apart from the time 'now', which is excluded from the meaning of the word 'then').
- The word 'we' might mean 'A and B' or it might mean 'A and B and Fred', depending on other facts in the common ground. If these facts do not make the meaning of 'we' unambiguous, B may have to use some longer descriptive phrase instead of 'we'.
- When A says 'Where to?', A and B both add mental state facts to the common ground: 'A does not know where we will set off to', and 'A wants to know where we will set off to' – because B's previous statement has made one place (where we will go to) the most prominent place in the common ground (other than the place denoted by the word 'here', which is prominent in the common ground, but cannot be the answer to A's question).

These examples bring out some points about the common ground (CG):

- The CG includes both physical facts and social/mental facts. It consists of facts, not words.
- The facts in the CG are fairly stable.
- The facts in the CG are ordered, in an order of **prominence**, or recency – so that whenever there is a choice between two facts, the speaker and the listener both choose the more prominent fact of the two.
- The normal course of a conversation is for each person to add facts to the CG, rather than remove facts or change them.
- The meaning of any word or phrase depends at least partly on facts in the CG, in a way that speaker and listener agree on (e.g. for any word or phrase, both choose the most prominent matching fact first).
- With each new utterance, the listener's version of the CG changes in the same way as the speaker's version, so as to keep the two versions of the CG in step.
- Conversation is in some sense a cooperative activity, where both parties try to keep their CGs in step, or initiate conversational repair if their CGs get out of step.
- Conversation does not require productive syntax or elaborate semantics. People can learn single word constructions to express specific meanings and make specific changes to the CG, before they learn many words to express complex meanings, or syntax to express meanings more productively (to express many meanings using few words).
- While early examples of language involve face-to-face conversations, a common ground is essential for any use of language (written, remote, etc.). Even in private verbal thought, one person will share a common ground with an imagined 'shadow audience' [Worden 2025].
- Because common ground always includes mental facts, use of a common ground always requires a mind-reading ability.
- People learn new words by hearing an unknown word in a situation where they know enough about the common ground to be able to infer what the word means, possibly in an over-specific manner, and then generalizing the meaning across several learning situations.
- The common ground in a conversation is a subset of each person's world model – a subset containing both physical and social/mental facts.

This is an informal description of the role of the common ground in language, which shows that common ground is central and essential to all language. Without a common ground, human language could not even get started.

There is a working computational model of language in the common ground [Worden 2025], which demonstrates all these points – how the common ground underpins the meanings of words, and how the speaker's and listener's versions of the common ground stay in step through all utterances, how people learn pragmatics, semantics and syntax, how pragmatics and semantics seamlessly overlap in any utterance, how mind-reading happens, and how diverse pragmatic phenomena such as presupposition, irony and metaphor are handled. This model uses the framework of Construction Grammars [Langacker 2008; Hoffman & Trousdale 2013], which has been developed over fifty years, in parallel with Chomskyan linguistics [e.g. Chomsky 1995] but is distinct from it.

The computational model of the Common Ground in [Worden 2025] is a Bayesian probabilistic model, and I note a few points about how it works:

- Facts in the common ground are represented by **Feature Structures**. Each feature structure is Directed Acyclic Graph (DAG) of a small number of nodes carrying slot values. Slots such as 'gender = male' typically constrain physical facts and other facts.
- Facts about mental states are represented by recursively nested feature structures.
- Each fact in the common ground has both a Bayesian probability, and a level of **prominence**, related to its recency (when it was added to the common ground).
- Peoples' learned knowledge of language (pragmatics, semantics and syntax) is represented by **constructions**. Constructions are feature structures with two branches or 'poles' – a phonetic pole and a semantic pole.
- Each word is a link between a phonetic pole (containing its sound) and a semantic pole (containing its meaning)
- The central operation in conversation is **unification** of feature structures.
- Unification is a Bayesian maximum likelihood way to combine two or more feature structures.
- To understand an utterance, in the context of the common ground, is to find a Bayesian maximum likelihood interpretation of it, by successive unifications (typically one unification per word). The common ground defines prior probabilities.
- There is another operation of **intersection** of feature structures, complementary to unification, which is used to learn new feature structures.
- The model of the common ground is implemented using conventional software engineering techniques, not as a neural net.
- There is a mathematical theory of the model, which can be used to prove important properties (such as that the speaker's and listener's versions of the Common Ground stay in step for each utterance)

Details are given in [Worden 2025]. Most of the techniques have been common parts of Cognitive Linguistics for many years [Hoffmann & Trousdale 2013].

This model of the common ground is an existence proof, that such models can be implemented, and can support a realistic model of actual conversations. Ideally, in the spirit of trying out alternative approaches, there would be several such models (for instance, built using Chomskyan linguistics rather than Cognitive Linguistics).

Language, using a common ground, is a central element of human social world; so if a world model is to include a model of the human social world (as it must), a theory of the common ground is a necessary part of it.

The Intentional stance is a consequence of the Common Ground. Whenever we converse with a person or device, we create a common ground in our minds – and within that common ground, infer that the person or device has mental states.

The main consequence of this section is that there is a requirement to implement a Common Ground as part of a World Model, and there are available techniques for doing so – without using neural nets.

## 9. Common Ground Failures and Hallucinations

It is evident from the previous section that LLMs do not have a common ground; they just appear to have one.

The main reason why LLMs have no common ground is that a common ground consists of facts, and LLMs do not have facts; they only have patterns of words. Another way to see the same point is that a common ground is a subset of a world model, and today's LLMs do not have a world model.

Why do LLMs appear to have a common ground? There are two main reasons:

1. LLMs have been trained on vast numbers of conversations and other language (mainly gathered from the Internet) in which there was a true common ground between people. The patterns of words which LLMs learn are patterns which emerge from true inter-personal common grounds. However, word patterns output by an LLM do not derive from a single, stable common ground; an LLM may easily switch common grounds from CG1 to CG2 in mid-utterance, if that produces an apparently acceptable pattern of words.
2. Because of our intentional stance, if anything can converse with us, we are instinctively inclined to believe that it has mental attributes, including a common ground which it shares with us. Even when there is no common ground, we invent it.

So LLMs do language without a common ground. Language without a common ground is at best a simulacrum of language, like a Hollywood film set. Because of our intentional stance, we accept the simulacrum.

The lack of a real common ground is the origin of many of the persistent failings of LLMs, such as hallucinations[3]. When we converse with a person about factual matters, our common ground includes a set of stable facts about the world: Latvia is a country, Leibniz died in the eighteenth century, and so on; and the common ground includes an expectation that 'we should stick to the truth'. When an LLM hallucinates, that is because it has no common ground that 'this is a factual discussion', 'I should only mention facts which are true' and so on; and it has no world model to tell it which facts are true (or indeed to tell it what 'true' means). An LLM hallucination is a particular kind of common ground failure. If the common ground included a fact that 'we are now telling fairy stories', hallucinations would not matter.

As another example of a common ground failure, consider playing chess against an LLM. An LLM can recite the rules of chess, but cannot complete a game obeying the rules. In a human game of chess, the common ground includes facts such as 'we are playing a game', 'each one of us is trying to win', 'we can only make moves that obey the rules of chess'. But an LLM has no such common ground. Therefore it is not capable of putting together its current proposed move with the rules of chess, and seeing that there is a conflict with a part of the common ground.

Even if an LLM could complete a game of chess according to the rules of chess, it would demonstrate a further limitation of pure LLM learning. The LLM has been trained on all the games of chess found on the Internet. These include vast numbers of games of an average or mediocre standard, as well as a few outstanding games. Having learnt mainly the patterns of mediocre chess, the LLM would play mediocre chess. The same goes for many other fields of endeavour, such as creative writing or science. An LLM could hardly distinguish astrology from astronomy; it can only repeat other peoples' assessments of them.

LLMs have a known tendency to reproduce the mediocre, which can be mitigated, as in Agent AIs, by post-training through Reinforcement Learning (RL). In this way, an AI system can be trained to seek the goal of winning the game, and thus to play better.

The intrinsic unreliability of LLMs arises from their lack of any true common ground, as well as their lack of world models. As any common ground is a subset of a world model, the two lacunae are closely related. To make LLMs reliable, we need to give them world models and a common ground capability.

## 10. A Possible Architecture for AI systems

How can we endow LLMs with world models and a common ground capability, to make them more reliable?

Suppose as part of an AI system architecture we can build a world model module and a common ground module. How should they be built?

The human common ground is essentially **reflective**; it can reflect on its own state and its own actions. For instance, to ask a simple question: 'What is X?', one must entertain reflective thoughts such as 'I do not know what X is", 'I want to know what X is' and 'the person I am conversing with may know what X is'. To have these thoughts, one must have reflective access to the contents of one's own mind.

Furthermore, the raw material for this reflection – facts like 'I do not know X' must be **stable**. The fact 'I do not know X' should not change, unless it is for a good reason, such as 'that person has just told me X'.

Suppose for a moment that the common ground module of an AI system was implemented as a neural net. Call this an NNCG. In section 3, we saw that the only possible engineering theory of a neural net is a complex, unstructured theory, consisting of vast numbers of anonymous facts such as 'the connection weight from unit 26437 to unit 67498 is 0.2569'. This theory of the NNCG does not lend itself to proving any useful general results about the NNCG - results such as: 'The NNCG is capable of a certain type of reflection about its own knowledge' or 'the results of NNCG reflection are stable against change'. Therefore an NNCG would not be capable of any useful or reliable reflection, and could not increase the reliability of an AI application.

It follows that the only way to implement a common ground module is to use the techniques of Symbolic AI, or equivalently, of conventional software engineering; so that the engineers who build the common ground module have a transparent, intellectually tractable theory of how it works – a theory that can be used to prove useful general properties, such as the stability of certain kinds of fact, or the capability of reflection.

This can be put in a different way by saying that an AI common ground needs to be implemented at David Marr's [1982] levels 1 and 2, of computational requirements, algorithms and data structures; not at Marr's level 3 of

---

[3] As pointed out by Seth [2025], an LLM strictly does not have hallucinations, because it has no internal experience which could be true or false to reality. Strictly speaking, LLMs do not hallucinate; they confabulate. But I shall continue to use the word 'hallucinate'

neural implementation. Neural nets exist only at Marr's Level 3.

The same argument implies that the World Model component of an AI architecture cannot be built as a neural net (a NNWM). There can be no tractable theory of a NNWM, which could be used to prove useful general properties (such as the stability of facts in the NNWM). The world model must be implemented using the techniques of conventional software engineering, at Marr's levels 1 and 2. Since the common ground is a subset of the world model, this close relation between the two implementations is not surprising. Both implementations require a property of **transparency**; and transparency is not a property of neural nets.

An architecture of an AI system interacting with a user is shown in figure 1.

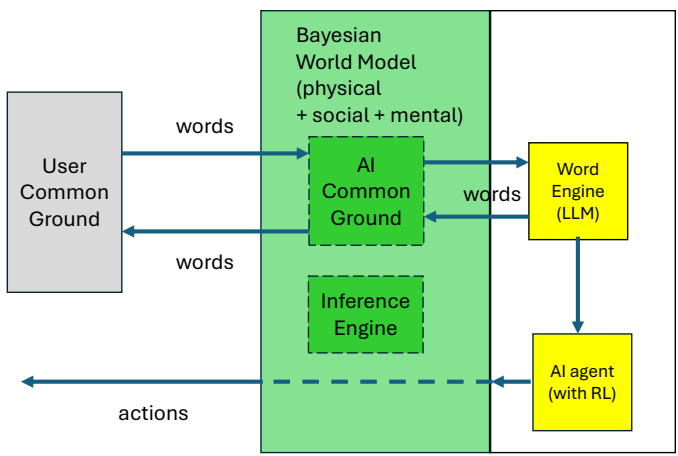

*Figure 1: Possible Architecture for an AI system. Components in yellow are implemented as neural nets; components in green are implemented by conventional software engineering, and there are tractable theories of those components. The inference engine has the functions proposed in [Bengio 2025b].*

The diagram shows a user interacting with an AI system, which uses one of today's LLMs. The user's common ground is in the mind of the user, as described in section 8; because of our intentional stance, this common ground is similar to a person's common ground in any human conversation. It consists of a set of facts about the physical world, the user's mental state, and what the user infers to be the state of the AI system.

The novel element in the architecture is the AI Common Ground, which sits between the user and the LLM, and controls the interactions between them, so that the user is shielded from any undesirable behaviour of the LLM (such as hallucinations or other common ground failures). This controlling protection property of the AI LLM must be provable in general cases from the theory of the AI common ground, and is tested as part of the certification of the AI system. For this, the AI LLM and the world model must be implemented by conventional, transparent software engineering techniques, so that the necessary theorems are provable and inspectable.

To do this, the AI Common Ground must be able to understand the words arriving at it from both sides – from the user and from the LLM – using conventional software engineering techniques. Here, to 'understand' is to convert words to a symbolic knowledge representation, with defined semantics. This is a challenging research problem at the frontier of natural language research; but it is not necessarily intractable. In the formalism of Construction Grammars [Huang 2017], the model of [Worden 2025] is a possible starting point; there may also a Chomskyan approach, although I have seen one being developed. All approaches should be tried, provided that they are transparent and tractable for engineering reasoning.

Some of the tasks of the AI common ground module are:

- To access and represent a large set of stable facts about the physical world, from the world model.
- To access and represent a large set of facts about the social and mental worlds
- To understand the words impinging on it from either the user or the LLM; that is, to convert those words into representations of physical, social and mental facts.
- To compare those representations with facts already represented in its common ground, and thus to incrementally build up a Bayesian maximum likelihood representation of the common ground, as a conversation proceeds – keeping its own common ground as much as possible in step with the user's common ground, or initiating conversational repair if needed.
- To compare its common ground with a set of possible common grounds which it has been certified to support; and if there is any violation, to intervene appropriately (e.g. to tell the user that the common ground is not supported; or to veto some words from the LLM to the user). It could, for instance, tell the user that certain conspiracy theories are not supported; or that it is restricted to a particular domain such as scientific research.

Building such an AI common ground module is a challenging research project; but it can be approached incrementally, building and slowly growing out its capability based on conservative principles: "If you don't understand it, don't let it happen", and "if anything might be unsafe, don't let it happen" - thus keeping both the user and the outputs of the LLM within the bounds of a provably safe capability. In this context, 'safe' means that the level of risk (quantified as a Bayesian probability, times a measure of severity) is below some certified acceptable level.

In an analogy, this architecture is like an automobile, with the LLM or AI Agent as the engine. The LLM is a utility for producing possible words, just as an engine converts gasoline to torque – and with the world model and common ground as the chassis, transmission, steering, brakes, and passenger compartment. The user value and safety of the system is provided by the AI common ground, and the LLM could be a replaceable commodity – useful even when it is unreliable.

Measures of safety of AI systems must be defined by society, in a way subject to testing, and proof from a theory of the AI device, so that the results of a limited number of test cases can confidently be extrapolated to general classes of use cases. These measures must include both a lack of short-term harm to individuals [Marcus 2020], and a lack of long-term harm to institutions such as the rule of law and a free press [Hartzog & Silbey 2026, Schroeder et al 2026]. The development and refinement of these measures will be an iterative process over several years – just as measures of safety for automobiles have been refined over many years.

## 11. Engineer AI

This section describes an alternative architecture for safe AI, which could be investigated in parallel with others.

[Bengio 2025b] has proposed the development of 'Scientist AI', an AI which does not act on the world, but which makes theories of the world based on evidence (I assume these include theories of the human social and mental world, as well as the natural world).

In section 2, I noted the close relationship between science and engineering. This means that if Bengio's Scientist AI is technically feasible, we can also create an Engineer AI, which is designed to follow the established engineering practice of section 2.

The only thing that Engineer AI is capable of doing is to create engineering devices, such as computer programs. It would follow established engineering practice, in that every time it designs an engineering device, it also creates a tractable engineering theory of the device. The theory (and testing) is used to prove safety results and certify the device for use.

The previous section described a 'runtime guardrail', in which the safety case for each AI application is re-made every time it is run. The approach of this section is more like a 'compile time guardrail' or 'build time guardrail' in which:

- Engineer AI builds a device to do some task, together with a tractable theory of the device.
- The theory and testing are used together to prove safety properties of the device
- The proofs are subjected to checking by non-AI methods

Any proof of safety properties depends on the world, so Engineer AI, like Scientist AI, depends on a world model. There are strong parallels between Scientist AI and Engineer AI, so that the challenges and difficulties for one are challenges for the other. I describe some of them.

Scientist AI includes both a world model and an inference engine, and Bengio's short-term plans for developing each of these use neural nets. Little is said in [Bengio 2025b] about a longer term development plan, but I suggest that the use of neural nets even in the short term is problematic, because it would imply that there is no tractable theory of the World Model or the Inference Engine – so there is no way to prove general results about them. These difficulties may be mitigated because:

- It is easier to verify a symbolic proof than it is to create one; verification of proofs can be done by reliable non-AI software engineering.
- It is easier to verify a Bayesian probabilistic inference than it is to create one. Verifying a chain of Bayesian inference requires integrating probability densities over many variables, which can be done by Monte Carlo integration, using a wide range of variable substitutions to make the integrations more efficient and to cross-check results. Integration accuracy increases with increasing computation, in an understood way.
- In a hybrid of the first two cases – a Bayesian probability distribution with some precise constraints, as for instance from geometry, kinematics or physics – the precise constraints can be represented as Dirac delta functions (infinitely narrow Gaussians) in the probability distributions. Integrals can still be done.

So if neural nets (or GFlowNets [Bengio et al 2021]) can generate candidate proofs or Bayesian inferences, the proofs and inferences can be verified reliably. They can be verified generally, or in specific cases which may be simpler and more illuminating, or in combination with testing.

There is also an issue of speed of learning. Scientist AI is intended to learn theories of the world, using model-based learning. [Bengio 2025b] writes: '*the bottleneck for information from the real world to train the Scientist AI is the length of the leading theories of the world model. We need* **about as much data as is sufficient to identify these theories**, *which we argue will be much less than the amount of data needed to directly train an inference machine from observed question-answer pairs.*'

This anticipates very efficient learning of Bayesian theories. There is a Bayesian theory of learning [Worden 2024, 2025], which implies that a probabilistic regularity of the environment can be learnt from very few examples – typically less than a dozen examples (enough examples, that the evidence for the regularity is statistically significant). This fast learning has been confirmed for associative conditioning in animals [Anderson 1990], and for language learning in humans [Bloom 1993, Worden 2021, 2025] where children can typically learn a word from as few as 3 or 4 examples of its use, in which they understand the common ground. This is no surprise, since animals need to learn very fast to survive. However, even after 50 years of development, neural net learning is many orders of magnitude slower than optimal Bayesian learning. It is not clear how this huge gap will be bridged in the next few years. Fortunately, there are computational models of Bayesian fast learning [Worden 2021], which can learn at Bayesian speeds.

Like Scientist AI, Engineer AI needs to include both a World Model (to prove that an engineering device is safe enough in the world), and a Common Ground (to carry on a dialogue with the user about the requirements for the proposed device). These should both use Bayesian probability, to avoid over-confident predictions.

As a simple illustration of how Engineer AI might work, I submitted the following request to various online AI interfaces:

> A circle in the (x,y) plane intersects the x axis at two points (x= a, y =0) and (x = b, y = 0). It intersects the y axis at two points, one of which is (x = 0, y = c), where a, b, and c are constants. Write a program to compute the area of the circle, and explain why it gives the correct answer.

The AI is required to develop a conventional software program, and a theory of how it works, sufficient to prove a key result – that it works correctly. If the AI succeeds in this, the program could be certified as safe for certain uses.

The problem is illustrated in figure 2:

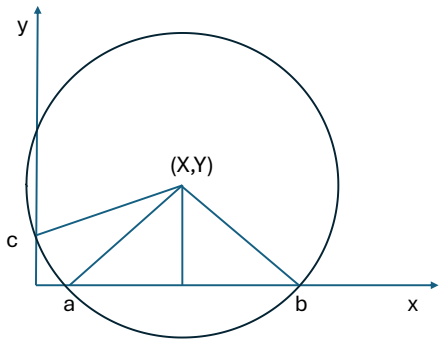

*Figure 2: Test problem used to illustrate the concept of Engineer AI.*

A successful solution would take a form such as:

Let the centre of the circle be the point (X,Y), and the radius be R.  X and Y obey
- $X = (a + b)/2$
- $R^2 = (Y - c)^2 + X^2 = Y^2 + (a - b)^2/4$

These are solved by
- $Y = (c^2 + ab)/(2c)$
- $A = \pi [Y^2 + (a - b)^2/4]$

(you can check these for special cases such as c = a,  c = b, or c tending to zero or infinity)

Most online AI services gave correct solutions to this problem, with a diverse set of answers (variable for each service), but they were not always correct. There were some incorrect programs, and sometimes the descriptions did not match the programs.  Today's AI is unable to reflect on its own problem-solving processes; even if it can do something correctly, it cannot reliably describe how it did it. The real AI process was '*my connection weights made me do it'*, and any other account of its process is factually incorrect. Some accounts were obviously incorrect.

However, all is not lost, as the AI often gets it right. A neural net can generate a candidate account of a process which might have given a correct program – leading to a candidate proof that the program is correct. The neural net can be asked to provide the proof in a formal language, such as mathematics. A non-AI proof checker can reliably test whether the proof is correct - and if it is not, keep rejecting it until the AI comes up with a correct proof. The same approach could work by checking any AI-generated Bayesian inference in Scientist AI or Engineer AI.

## 12. Conclusions

This paper has examined the underlying reasons why applications of neural nets, including LLMs and Agentic AI, are unreliable.  The core reason is that neural nets depart from an established engineering practice of more than 200 years, that engineers understand the devices they build, and can reason about them. Before neural nets, engineers had tractable theories of the devices they built. These theories allow them to prove useful general properties of the devices, and to extrapolate from limited numbers of test cases to unlimited numbers of use cases. With neural nets, there is no tractable theory of the device, and it is no longer possible to extrapolate from test results. So neural nets are intrinsically unreliable.

To apply neural nets reliably, the neural net must be enclosed in a controlling architecture, with the controlling component being the subject of a tractable theory, so that the whole device can be proved to have useful and safe properties, even when the neural net itself does not.

As part of the controlling architecture, it has been proposed [Marcus 2020, Bengio 2025b] that AI systems should include a stable world model. I argue that the world model needs to include not only a model of the physical world, but also models of the social and mental worlds. The last two models present significant unsolved research challenges.

To analyse how to address these challenges, I discuss the nature of human language, and its dependence on a Common Ground of understanding of the current situation, shared between participants in a conversation. The common ground includes physical, social and mental facts, and is a subset of a world model. It is a vital part of all human language. There is a working computational model of how the common ground operates in the human mind, but more research is needed. Because of the central role of the common ground in language and human nature, a theory of the common ground is a key element of a theory of human nature and mental life, as required in an AI world model.

To build reliable AI applications, it is necessary to implement both a World Model component, and a Common Ground component, using conventional software engineering techniques, so that there are tractable engineering theories of those components.

Another possible architecture is to build an 'Engineer AI' which designs engineering devices to complete agent tasks, together with a theory of the device which can be used to prove it is safe. This applies the guardrail to AI at build time rather than at run time, and still depends on a Bayesian world model to prove that the device is sufficiently safe in the world.

These are a possible vision for the application of AI which is not prone to the endemic risks of today's AI systems. How realistic is that vision, in societal and commercial terms?

The vision requires governments define safety criteria for AI systems, and require certification of AI systems against those criteria. This raises difficult questions, not least of trans-national coordination, and the protection of AI systems against bad actors such as rogue states which conduct cyber-warfare using AI. Success is by no means guaranteed – as is shown by the recent unanticipated effects of social media - but it is certainly better for governments to try, than just to give up. The existence of taxation shows that it is possible for governments of many nations, in spite of their profound differences, to exercise some unified control over the large commercial organisations that are likely to be the creators of AI systems. One should not underestimate the role of individual choice – for instance, people learning that properly controlled AI systems work better for them than bare uncontrolled LLMs, and choosing certified AI systems in the market.

Finally, a plea for urgency. Scientific research has typically proceeded at an academic pace, with major advances arriving unpredictably. The problem of AI safety is an urgent one for society, and we cannot afford to wait for breakthroughs. Government and commerce needs to make AI safety the very first priority for research - not only to avert the dangers of AI, but also to enable AI to contribute fully to solving other existential problems of society, such as climate change.